\documentclass[12pt]{article}
\usepackage{epsfig}
\usepackage{epsfig}
\usepackage{amssymb}
\usepackage{amsfonts}

\usepackage{amsmath}
\newskip\humongous \humongous=0pt plus 1000pt minus 1000pt

\newif\ifdtup

\catcode`\@=11

\@addtoreset{equation}{section}
\def\theequation{\thesection\arabic{equation}}

\def\@normalsize{\@setsize\normalsize{15pt}\xiipt\@xiipt
\abovedisplayskip 14pt plus3pt minus3pt%
\belowdisplayskip \abovedisplayskip
\abovedisplayshortskip \z@ plus3pt%
\belowdisplayshortskip 7pt plus3.5pt minus0pt}

\def\small{\@setsize\small{13.6pt}\xipt\@xipt
\abovedisplayskip 13pt plus3pt minus3pt%
\belowdisplayskip \abovedisplayskip
\abovedisplayshortskip \z@ plus3pt%
\belowdisplayshortskip 7pt plus3.5pt minus0pt
\def\@listi{\parsep 4.5pt plus 2pt minus 1pt
\itemsep \parsep
\topsep 9pt plus 3pt minus 3pt}}

\relax

\catcode`@=12

\catcode`\@=11

\def\section{\@startsection{section}{1}{\z@}{3.5ex plus 1ex minus
.2ex}{2.3ex plus .2ex}{\large\bf}}

\def\thesection{\arabic{section}.}

\def\appendix{\setcounter{section}{0}
\def\thesection{Appendix \Alph{section}:}
\def\theequation{\Alph{section}.\arabic{equation}}}

\newcommand{\be}{\begin{equation}}
\newcommand{\ee}{\end{equation}}
\newcommand{\bqa}{\begin{eqnarray}}
\newcommand{\eea}{\end{eqnarray}}
\newcommand{\beas}{\begin{eqnarray*}}
\newcommand{\eeas}{\end{eqnarray*}}

\newcommand{\bquo}{\begin{quote}}
\newcommand{\enqu}{\end{quote}}

\def\Tr{ \hbox{\rm Tr}}
\def\const{\hbox {\rm const.}}
\def\o{\over}

\def\brc{\langle}
\def\ckt{\rangle}

\def\Im{\hbox {\rm Im}}
\def\diag{\hbox{\rm diag}}

\def\si{\sigma}
\def\tr{\hbox {\rm Tr}}

\begin{document}

\begin{titlepage}
{\hfill      IFUP-TH/2005-32  }
\bigskip
\bigskip


\begin{center}
{\large  {\bf
Nonabelian Confinement near Nontrivial Conformal Vacua
 } }
\end{center}


\renewcommand{\thefootnote}{\fnsymbol{footnote}}
\bigskip
\begin{center}
{\large
 Kenichi KONISHI $^{(1,2)}$,  \\  Giacomo
 MARMORINI $^{(3,2)}$,
 Naoto YOKOI $^{(2,1)}$}
 \vskip 0.10cm

\end{center}

\begin{center}
{\it      \footnotesize
Dipartimento di Fisica ``E. Fermi" -- Universit\`a di Pisa $^{(1)}$, \\
Istituto Nazionale di Fisica Nucleare -- Sezione di Pisa $^{(2)}$, \\
Largo Pontecorvo, 3, Ed. C, 56127 Pisa,  Italy \\
Scuola Normale Superiore - Pisa $^{(3)}$,
 Piazza dei Cavalieri 7, Pisa, Italy } \\
 
 \smallskip
 
 { \small  \texttt    konishi(at)df.unipi.it, g.marmorini(at)sns.it, yokoi(at)df.unipi.it }

\end {center}

\noindent
\begin{center} {\bf Abstract} \end{center}

{We discuss some aspects of confinement and dynamical symmetry breaking  in the so-called nonabelian Argyres-Douglas vacua, which occur very generally  in  supersymmetric theories.
These systems are characterized by strongly-coupled nonabelian monopoles and dyons;  confinement and dynamical symmetry breaking
are caused by the condensation of monopole composites, rather than by condensation of single weakly-coupled monopoles.
In general, there are strong constraints on
which kind of monopoles can appear as the infrared degrees of freedom,
related to the proper realization of the global
symmetry of the theory. Drawing analogies to some of the phenomena found here,   we make a  speculation  on the ground state of  the standard QCD.
  }

\vfill

\begin{flushright}
\today
\end{flushright}
\end{titlepage}

\section{Introduction}

Our understanding of the quantum  behavior of nonabelian monopoles  in  4D gauge theories \cite{NAmonop}-\cite{ABEKM} has greatly improved in the last few years.  By exploiting the knowledge of the exact solutions in theories with ${\cal N}=1$  or ${\cal N}=2$ supersymmetry \cite{SW1}-\cite{CDSW} we know for instance that there are systems in which the 't Hooft-Mandelstam
scenario of confinement  (dual Meissner effect) \cite{TM}   is  indeed
realized dynamically.  However, in all such systems confinement is accompanied by dynamical abelianization, with characteristic enrichment of Regge trajectories, a feature which is not shared by quantum chromodynamics (QCD) \cite{Yung,Strass}.

    Actually, a more general class of supersymmetric theories  with massless quark fields   exhibit a rather different picture of  confinement and dynamical symmetry breaking.  First of all,  confinement is typically  described as a dual Meissner effect,  of a {\it nonabelian kind},  where  both condensing monopoles and confining strings carry unbroken nonabelian fluxes.
  The properties of these nonabelian superconductors \cite{HT,ABEKY}  are being intensely investigated \cite{NAVortex}.


     These   nonabelian confining vacua studied so far,  can be further classified, roughly speaking,
into two subclasses  of systems; the first one characterized  by weakly coupled monopoles,
and the second  involving relatively non-local set of strongly-coupled  monopoles and dyons.   The first class of systems can be described by a local (dual) effective Lagrangian, which can be easily and exactly analysed, for instance,  by using the Seiberg-Witten solutions;  confinement and dynamical symmetry breaking are explicitly described in terms of the nonabelian monopoles and vortices. From the point of view of classification of conformal theories, these correspond to trivial (infrared-free) conformal theories, perturbed by a ${\cal N}=1$ potentials. 
   Although interesting,  this first group of  systems  (typical examples are the  ``$r$ vacua''  of  ${\cal N}=2$, $SU(N)$  SQCD) cannot be considered as a good model for QCD,  either.  For instance, the problem raised in \cite{Yung} of non linear Regge trajectories would persist also in these cases.


In the present  paper,  we concentrate our attention on  the second
class of nonabelian vacua,  where  the infrared degrees of freedom
involve relatively non local and  in general strongly-coupled dyons.
The theory is close to  ({\it i.e.}  a perturbation of)  a
non-trivial  fixed-point theory, which is a superconformal field theory
(SCFT). These systems unfortunately defy the traditional
effective-Lagrangian approach, and as a result it is much more
difficult to understand what is happening in the infrared.
Nevertheless, by fully mobilizing our general knowledge such as the
Nambu-Goldstone theorem, Seiberg-Witten curves, instanton-induced
effective action, holomorphic properties due to supersymmetry, vacuum
counting, decoupling theorem, exact anomalous and non-anomalous
Ward-Takahashi identities, universality of conformal theories,
Seiberg's duality, some new  results on ${\cal N}=1$ susy gauge
theories \cite{DV,CDSW}, and so on, one can get a fairly precise
picture of physics in these vacua.   It is the purpose of this paper
to discuss various aspects of these confining vacua, characterized
by  highly non-trivial monopole dynamics.

\section{Physics of nonabelian Argyres-Douglas  vacua:\\  confinement versus  dynamical symmetry breaking}

 Study of nonabelian generalization of electromagnetic duality  goes back to work
 even before the seventies \cite{NAmonop}-\cite{GNO}. The idea that such duality
 is a {\it symmetry} of the system (Montonen and Olive \cite{MO}), is believed to hold in superconformal ${\cal N}=4$ gauge theories. In the context of
 asymptotically free gauge theories, the first example  of nonabelian duality was found  by Seiberg \cite{Sei} and others \cite{KS},
in the context of  ${\cal N}=1$  theories.  In the supersymmetric
QCD  (SQCD), with the number of flavor  in the so-called conformal
window, $\frac{3 \, N_{c}} {2}  < N_{f}  <  3\, N_{c}$,  the theory
at the origin of the vacuum moduli space (with all scalar VEVs
vanishing)  flows into an infrared fixed point:   the theory becomes
superconformal (SCFT).  This theory is described either by the
original $SU(N)$ QCD, or by the dual $SU(N_{f}-N)$ theory with
$N_{f}$  dual quarks.  A review on various dualities and their
relations can be found in \cite{Strass}.

Somewhat analogous  superconformal theories,   appearing as an
infrared fixed-point theory,  were  discovered by  Argyres and
Douglas \cite{AD} and others \cite{SCFT,BF,Eguchi},  in the context of
 ${\cal N}=2$  supersymmetric gauge theories.   The example studied
in \cite{AD}  is a pure ${\cal N}=2$  supersymmetric $SU(3)$
Yang-Mills theory:  the infrared theory in question is an effective
$U(1) \times U(1)$  theory, where one of the $U(1)$ factor is
strongly coupled. Interest in this kind of systems  (which we shall
call Argyres-Douglas vacua in this paper) arises from the fact that,
in contrast to the cases discussed by Seiberg and others, the nature
of the low-energy degrees of freedom is  better understood and, in
particular, because they are known to include a mutually nonlocal set
of dyon fields.  In the specific case of  the interacting $U(1)$
SCFT of Argyres and Douglas,   the IR  ``matter'' degrees of freedom
are a magnetic monopole, a dyon and an electron. The beta function
cancel among the contributions from these relatively nonlocal fields
\cite{AD,SCFT}.   See also \cite{KY} for a further discussion on the renormalization group flow near such fixed points.

We are interested here in nonabelian analogues of the Argyres-Douglas vacuum,  appearing in various theories.

\subsection{Argyres-Douglas  vacua of   $SU(N)$  supersymmetric QCD} \label{SUNSCFT}

In the ${\cal N}=2$  $SU(N)$  SQCD, the degrees of freedom in UV are
(in ${\cal N}=1$ formalism) the gauge multiplet $W= (A_{\mu},
\lambda)$, the adjoint chiral multiplet $\Phi= (\phi, \psi)$,  and
the quark multiplets $Q_{i}= (q, \psi_{Q})_{i}$ and ${\tilde Q}_{i}=
({\tilde q}, {\tilde \psi}_{Q})_{i}$, in the fundamental
(antifundamental) representation of the gauge group ($i=1,2,
\ldots, N_{f}$). We start with small quark masses $m_{i}$ and we
break the ${\cal N}=2$ supersymmetry to ${\cal N}=1$ by a small
superpotential (mass term) for the adjoint chiral multiplet,
\be \mu\,\tr \, \Phi^{2}|_{F} = \mu  \, \psi \psi + \ldots.
\ee
This theory has a large vacuum degeneracy (vacuum moduli):  the
vacuum we are interested classically corresponds to the one
characterized by the scalar VEV,
\be   \brc {\phi} \ckt  = \frac{1}{\sqrt 2}  {\rm diag} \, (-m_{1}, -m_{2}, \ldots, -m_{r},
c,c,\ldots, c), \qquad c = \frac {1}{N_{c}-r} \sum_{k=1}^{r} m_{k},
\label{rvacua}\ee
\be \brc {q}_{i}^{\alpha} \ckt   =
\delta_{i}^{\alpha} \,  \sqrt {\mu \, m_{i}}, \qquad i=1,2, \ldots,
r, \label{globalSB}\ee
that is, with the gauge group classically
broken to
\be  SU(N_{c}) \to  \prod_{i} U_{i}(1)  \times SU(N_{c}-r)
\stackrel{ m_i \to m }{\longrightarrow} U(r) \times SU(N_{c}-r).
\label{clasbr} \ee
The $SU(N_{c}-r)$ sector becomes strongly
interacting in the infrared and breaks itself to the maximally
abelian subgroup \cite{DS}.  The $\prod_{i} U_{i}(1) $ factor group
in Eq.~(\ref{clasbr}), on the other hand, gets enhanced into the
$U(r)$  group in the equal mass (and/or massless)   limit $m_{i} \to
m $   (or $m_{i }\to 0$). This $SU(r)$  is infrared free if  $r <
\frac {N_{f}}{2}$, as is seen from the effective quark mass terms
\be     Q_{i}  \, ( \brc \sqrt{2} \Phi \ckt  +  m_{i} ) \, {\tilde
Q}_{i},
 \ee
 and Eq.~(\ref{rvacua}).

 The critical cases  $r= \frac {N_{f}}{2}$,  where
the theory becomes conformal invariant  at low-energies, are  our main interest in this section.

The simplest such example appears in the  $SU(3)$ gauge theory with
$N_{f}=4$ flavors. In the Argyres-Douglas vacuum of this theory, the
low-energy effective gauge group is  $SU(2) \times U(1)$. If the
masses
   $m$ are large compared to the dynamical scale of the theory $\Lambda$, the theory is basically the local
   $SU(2)$  theory with $N_{f}=4$ flavors, which as is well known, is conformally invariant, with $\beta=0$.

As $m \to 0$, however, the flow of the theory into the infrared
fixed point occurs in a nontrivial way. The low-energy physics of
${\cal N}=1$ vacua are encoded in the structure of the singularities
of the Seiberg-Witten curve \cite{curves}, which in this case reads,
\be y^2 =  \prod_{a=1}^{3}  (x-\phi_a)^2 -  \prod_{ i=1}^4    (x+m_i ) \equiv    (x^3  -   U  x  -
V )^2  -  4 \Lambda^{2} \prod_{ i=1}^4    (x+m_i ), \label{SWbis}
\ee
where $U={1\over 2}\brc \tr \Phi^{2} \ckt$ and $V={1\over 3} \brc
\tr \Phi^{3} \ckt $ parametrize inequivalent vacua
\footnote{Eq.(\ref{SWbis}) represents a one dimensional complex
surface (curve), which can be thought as a hypertorus with genus
two.   The low-energy effective coupling constant and  $\theta$
parameter, as well as the masses of the short multiplets (BPS
states) are expressed as integrals of certain differential forms along the nontrivial cycles on
this hypertorus, which  constitutes
the Seiberg-Witten solution \cite{SW1}-\cite{AS}.  Curves    such as
(\ref{SWbis}) compactly encode all the perturbative and
nonperturbative  effects.
   }.
For {\it equal bare quark masses}    ($m_i=m$),   it simplifies:
\be  y^2 =  \prod_{a=1}^{3}  (x-\phi_a)^2 -    (x+m )^4    \equiv    (x^3  -   U  x  -
V )^2  -  4 \Lambda^{2} (x+m )^4.
\label{SW} \ee
The  $r=2$  (Argyres-Douglas)  vacuum corresponds to  the point,  $\diag\,  \phi =  (-m, -m,  2m )$,
{\it i.e.},
\be    U = U^{*}=   { 3 m^2}; \qquad   V = V^{*} = 2 m^3,
\label{Sing}\ee
 where the curve exhibits  a singular behavior  (the bi-torus degenerates into a sphere),
\be   y^2 \propto   (x+m)^4
\ee
corresponding to the unbroken $SU(2)$  symmetry \footnote{ The singularity (\ref{Sing})  splits into six separate nearby singularities when the quark masses are
taken to be slightly unequal and generic (the sextet vacua).}.

In order to find out the nature of the infrared degrees of freedom in the SCFT  vacuum (\ref{Sing})
one must determine  the  loci  in the $(U, V)$  space  near   $(U^{*}, V^{*})$, along which some massless particles are present
(corresponding to a double branch point of the curve Eq. (\ref {SW})), and study how various quantities transform as
one goes around such loci (monodromy matrices).
This problem has been analyzed in detail in \cite {AGK};   the low-energy degrees of freedom are found to carry  the magnetic and electric  $U(1) \times U(1)$  charges,  shown in  Table \ref{monopcharges},  with the first $U(1)$ factor
(magnetic or electric) referring to the subgroup of the $SU(2)$.
The system having ${\cal N}=2$ supersymmetry, there are also particles ${\tilde M}^{\alpha}$,  ${\tilde D}^{\alpha}$,  ${\tilde E}^{\alpha}$, with conjugate  gauge quantum numbers.
\begin{table}[h!]
\begin{center}
\begin{tabular}  {|l|l|} \hline Particles  & $(g_{1},g_{2}; q_{1}, q_{2})$ \\
\hline
$M_{1}, {M}_{2}$ & $(\pm 1,1; 0,0)^4$ \\ $D_{1},{D}_{2}$ & $(\pm 2,-2;\pm 1,0)$  \\
$E_{1}, E_{2} $&$(0,2; \pm 1,0)$    \\
\hline
\end{tabular}
\caption{\footnotesize The charges of the massless doublets.   $g_{i}$ ($q_{i}$)  is  the magnetic (electric) charge with respect to the $i$-th $U(1)$ factor.}
\label{monopcharges}
\end{center}
\end{table}

 The superscript in the table indicates the multiplicity of the massless particle present.
 The pair of particles carrying opposite charges with respect to the first $U(1)$  (magnetic or electric)  factor,  can be interpreted naturally as forming a
doublet of the $SU(2)$.  This way we arrive at the conclusion that
 there are  massless monopole doublets carrying the ${\underline 4}$ flavor charge of $SU(4)$, and a dyonic  and an electric doublets which are singlets of the global  $SU(4)$.    The particles in the table carry indeed relatively nonlocal charges, {\it i.e.}, nonvanishing relative Dirac unit \cite{TM},
 \be      \sum_{i=1}^{2}    (  g_{i} q_{i}^{\prime} -    g_{i}^{\prime} q_{i} )   \ne 0,  \quad  {\rm Mod} \,\,[ N ]
 \ee
(for $SU(N)$, here $N=3$),    and the theory is superconformal \cite{SCFT}.
Let us recall that the cancellation of the beta function has been checked in \cite{AGK} by generalizing the 
argument by Argyres and Douglas \cite{AD}, to our nonabelian SCFT.  Here, in contrast to the case studied in \cite{AD}, the gauge multiplet contributes. In the dual base in which $M, {\tilde M}$ fields are coupled minimally to the dual gauge bosons, the contribution of four flavors of  $M, {\tilde M}$  cancel  that of the $SU(2)$ gauge multiplet.
The contribution of $D$ and $E$  fields to the beta function must be computed in the base where these are coupled
locally to the appropriate (dyonic or original) gauge bosons,  then the result transformed back to the magnetic base. 
It turns out they cancel precisely 
\be
1 + (2\, \tau^{*}  +1)^2=0,  \ee 
at the critical coupling constant, $ \tau^*=\frac{-1+i}{2}$,
the value found independently from the study of the shape of the hypertorus in the SCFT limit \cite{AGK}. 

The $\mu \, \Phi^{2}$ perturbation breaks superconformal invariance, and the theory confines.

This precise knowledge on the infrared degrees of freedom from the Seiberg-Witten curve can be  combined with the pattern of the symmetry breaking following from the independent analysis made at large $\mu$. Due to the holomorphic dependence of physics on the parameter $\mu$ there cannot be any phase transition at a finite value of  $|\mu|$.  At large $\mu$,  (where the nonperturbative dynamics is that of ${\cal N}=1$ theory)  the instanton-induced superpotential is known from the earlier studies, and one can easily determine the symmetry breaking pattern in this vacuum \cite{CKM},
\be   SU(4) \times U(1) \to U(2) \times U(2).
 \label{strong} \ee
  One is then  led  to  conclude
 that the symmetry breaking at small $\mu$   is caused by the bilinear condensate  \footnote{ $\alpha, \beta = 1, 2$ are dual color indices, $i,j =1,\ldots, 4$
 are the flavor indices of $G_{F}= SU(4)$.}
 \be   \brc  \epsilon_{\alpha \beta} \,  M_{i}^{\alpha} \, M_{j}^{\beta}\ckt \ne 0, \qquad
 \brc  \epsilon^{\alpha \beta}  \,  {\tilde M}^{i}_{\alpha} \, {\tilde M}^{j}_{\beta}\ckt \ne 0,
 \label{condens}
 \ee
due to the strong magnetic $SU(2)$ interactions.  Note that the condensate (\ref{condens}) is antisymmetric in flavor and, as required,  reproduces the correct symmetry breaking pattern, (\ref{strong}).

   A crucial observation made in \cite{AGK}  is the following.  As the bare quark masses $m_{i}$ are taken slightly different from each other,
   the vacuum we are considering splits into six nearby vacua. Each of them is a {\it local}  Abelian $U(1)\times U(1)$ theory, with a pair of massless monopoles,
   which condense upon ${\cal N}=1$ perturbation.
 The problem is that  the condensates of   the   abelian monopoles
 \be    \brc M \ckt =  \const \, \sqrt {\mu \Lambda}
 \ee
in each vacuum,  in fact, vanish  ($\const \to 0$)    in the limit  $m_{i} \to m$   we are interested in.\footnote{ Analogous  phenomenon has been noted in \cite{GVY}.}   Dual  superconductor picture of confinement \`a la 't Hooft-Mandelstam \cite{TM} (with $U(1)^{2}$ gauge symmetry)    is, therefore,   not the correct mechanism of confinement in  the present system.

We believe that here the confinement (and the symmetry breaking) is caused by the strong interaction effects among the nonabelian monopoles, (\ref{condens}),   which are  entirely missed  in the abelian local effective-action description at  each of the six vacua, formally valid before the SCFT limit is taken.  Actually, the validity (in the mass scale)  of the abelian local effective  action of each vacuum,    is limited from above  by the masses of the massive nonabelian gauge bosons, which tend to zero in the $m_{i} \to m$ limit. In order to  correctly understand the infrared physics, one must take into account the full $SU(2)$  gauge interactions.

In more general
$r=\frac{N_{f}}{2}$  vacua  of $SU(N)$ theory, one expects again that nonabelian monopoles $ M_{i}^{\alpha}$  in the fundamental representation of $SU(\frac{N_{f}}{2})$  gauge group
and in the fundamental representation of the global $SU(N_{f})$ group,  will be responsible for confinement/dynamical symmetry breaking. From  the known symmetry breaking pattern \cite{CKM}
\be    SU(N_{f}) \times U(1) \to   U(\frac{N_{f}}{2}) \times U(\frac{N_{f}}{2}), \label{SB}
\ee
we conclude that the baryonlike condensate
 \be   \brc  \epsilon_{\alpha_{1} \alpha_{2} \ldots \alpha_{N_{f}/2}}\,    M_{i_{1}}^{\alpha_{1}} \, M_{i_{2}}^{\alpha_{2}} \, \ldots M_{i_{N_{f}/2}}^{\alpha_{N_{f}/2}} \,\ckt \ne 0, \label{condensgen}
 \ee
forms in this vacuum.
The mesonlike condensates
\be    \brc  M_{i}^{\alpha}  \, {\tilde M}_{\alpha}^{j} \ckt    = \const \, \delta_{i}^{j},
\ee
might also form, but they do not modify the symmetry breaking pattern (\ref{condensgen}), being singlet of
$G_{F}= SU(N_{f})  \times U(1)$.

\smallskip

\noindent{\bf Baryon number conservation}

There is, actually,  one subtle issue in Eq.(\ref{condens}) and Eq.(\ref{condensgen}) concerning   the
quark (baryon) number conservation.   The analysis   made  at large $\mu$  (where the order parameter of symmetry breaking is the meson  condensates  $\sim \brc Q \, {\tilde Q}\ckt $    only)  shows that the baryon number $U(1)$  is unbroken, while the nonabelian part $SU(N_{f})$ is broken,  as in Eq.(\ref{strong}) and Eq.(\ref{SB}).
On the other hand, the infrared degrees of freedom of the theory at small $\mu$  are the nonabelian monopoles, carrying in general nontrivial flavor quantum numbers, as  ${\underline {4}}$  of $M, {\tilde M}$ of Table \ref{monopcharges} .
Since  the soliton monopoles usually carry also fractional $U(1)$ charges - unless forbidden by a symmetry - one might wonder whether
our picture of the infrared physics  based on strongly interacting monopoles is consistent.

It is a highly nontrivial check of consistency of our claim,   that  the dual quarks  of the quantum $r$ vacua ($r \le \frac{N_{f}}{2}$)
carry  exactly vanishing baryon number \cite{APS,CKM}.    The phenomenon of the quark-number quenching due to quantum effects
of the {\it massless}  monopoles,  has been further clarified  in \cite{CKT}.

\subsection{ $USp(4)$, $N_{f}=4$ }

Another simple but very instructive  example of nonabelian Argyres-Douglas vacua occurs in $USp(4)$  theory with $N_{f}=4$  flavors.  This system
has been analyzed by Auzzi and Grena \cite{RR}.     A characteristic feature of this model, which makes it really interesting,  as compared to the $SU(N)$ theories discussed above,   is the fact that   it possesses
a nontrivial chiral symmetry,  $G_{F}= SO(8)$ \footnote{In a nonsupersymmetric version of the model,  the global symmetry would be
$SU(2 \, N_{f})$,  but the Yukawa interaction typical of ${\cal N}=2$  supersymmetry reduces the symmetry to   $SO(2N_{f}) \subset SU(2N_{f})$.}.
The action of this theory is the standard ${\cal N}=2 $ action    with
 superpotential,
\be
W= { 1 \o \sqrt2} \, Q_a^i \Phi_b^a Q_c^i \, J^{bc} +
{m_{ij} \o 2} Q_a^i Q_b^j \, J^{ab} \, ,\qquad i,j =1,2,\ldots 2  N_{f},
\label{uspsuperpot}
\ee
where $J = i\si_2 \otimes {{\bf 1}}_{N}$   and
\be
m = - i \si_2 \otimes \diag \, (m_1, m_2, \ldots, m_{N_{f}}) \, .
\ee
In the $m_i \to 0$ and $\mu \to 0$  limit, the global symmetry is
$SO(2 \, N_f) \times Z_{2 N +2 -  N_{f} } \times  SU_R(2)$.

This theory ($N_{c}=2, \, N_{f}=4$)   has seventeen vacua (for unequal  $m_{i}$ and for $\mu \ne 0$). These  degenerate into two  ``Chebyshev''  vacua\footnote{These vacua are characterized by the point of the moduli $\{\phi_{i}\} $, which can be  determined by use of some particular properties of the Chebyshev polynomials - trick first used by
Douglas and Shenker \cite{DS}. }   and one ``special'' vacuum,   in the limit, $m_{i} \to 0$.   The Chebyshev vacua, which are the new  SCFT,  are the system of our interest here.
Upon ${\cal N}=1 $  perturbation
\be \Delta W  =  \mu \, \Tr \, \Phi^2
\ee
it can be shown that the special vacuum remains unconfined (free magnetic phase) while the two Chebyshev  vacua  are in confinement phase.
To study the infrared properties of these vacua it is necessary to determine the light  degrees of freedom.

This system has been  carefully studied,  by following the method used for the previous case \cite{AGK},  by Auzzi and Grena \cite{RR}.   Their result is summarized in the following table.  They find that,  as compared to the $SU(3)$ SCFT  considered above, there is one extra doublet in this system  ($C_{1}$, $C_{2}$ in Table.~\ref{monopcharges2}).     The structure of  the singularities (loci in the quantum moduli
space where some dyon becomes massless) and the monodromies around each part of the singular loci, hence the charge determination of the Table.~\ref{monopcharges2},  have been double-checked independently  by the present authors.

\begin{table}[h]
\begin{center}
\begin{tabular}  {|l|l|} \hline Particles  & Charge \\
\hline
$M_{1}, M_{2}$ & $(\pm 1,1,0,0)^4$ \\ $D_{1},D_{2}$ & $(\pm 2,-2,\pm 1,0)$  \\
$E_{1}, E_{2}$&$(0,2,\pm 1,0)$    \\
$C_{1}, C_{2}$&$(\pm 2, 0,\pm 1,0)$    \\
\hline
\end{tabular}
\caption{\footnotesize The charges of the massless doublets in one of the SCFT vacua. }
\label{monopcharges2}
\end{center}
\end{table}

Given the charges of the massless particles and given the symmetry breaking pattern,
\be  SO(8)  \to   U(4),
\ee
(known from the analysis at large $\mu$ \cite{CKM}),  we are forced to conclude that
the monopole pair $M, {\tilde M}$ condenses as
\be   \brc M_{a}^{i}   {\tilde M}_{j}^{a} \ckt  = v\,  \delta_{j}^{i} \ne 0,  \qquad  a,b = 1,2 \quad {\rm and} \quad i,j =1,\ldots,4. \label{monocond}
\ee
Although the set of massless particles are rather similar to those found in the $r=2$ vacuum of the  $SU(3)$  theory, we do not expect ``baryonlike'' condensate
(\ref{condens}) to  form in  this  system.
Other condensates such as 
\be  \brc C_{a}   {\tilde D}^{a} \ckt,   \qquad    \brc D_{a}   {\tilde C}^{a} \ckt,  \qquad \brc C_{a}   {\tilde C}^{a} \ckt,
\ee
etc., including the new doublet $C$,  might well form, but would not modify the symmetry breaking pattern.

The difference in the massless spectrum and in the dynamics of this system, as compared to those in the $SU(3)$  theory discussed in the previous section, can be attributed to the fact  that  these two SCFT's  belong to two different universality classes.
See below for more about this point.

\subsection {Light nonabelian monopoles in $USp(2N)$  theories \label{lightmpl}}

It has been  shown  \cite{CKM}  that {\it all}  of  the  confining  vacua of $USp(2N)$  theories with $m_{i} \to 0$, $\mu \ne 0$,  with $N_{f}\ne 0$ flavors,
are perturbation of  nontrivial superconformal vacua, as in  the $USp(4)$ example discussed above.  (The same  holds for  $SO(N)$  gauge theories as shown in \cite{CKKM}).  The infrared physics is not described by a local effective Lagrangian
and is difficult  to analyze.  However,  the chiral symmetry breaking pattern
\be  SO(2N_{f}) \to  U(N_{f}) \label{similar}
\ee
is  {\it known}  from the behavior of the system at large $\mu$ \cite{CKM}.
 There is actually some intriguing similarity  \cite{CKM}  between (\ref{similar}) and  the symmetry breaking pattern   in the  real-world  QCD,  (\ref{SBQCD}), so it is   important to attempt to understand better the physics of this system.

It is possible that condensate analogous to (\ref{monocond}) forms in genereal $USp(2N)$ theory,  but what are the global quantum numbers carried by the monopoles?  Which kind of monopoles are present?  How do they interact?

In the absence of the local effective action valid at low energies it is not an easy task to answer such questions.  To work out the
singularity structures and the monodromy matrices around each subsingular loci, as has been done in the simplest cases
\cite{AGK} and \cite{RR}, seems to be out of question.
Fortunately, one can vary certain parameters in the system and go to the regimes where physical interpretation is easier.

In particular, it is useful to vary the quark masses, $m_{i}$,
although we are really interested in what happens in the $m_{i }
 \to 0$ limit.
  First, let us choose nonvanishing but equal masses,  {\it i.e.},  $m_{i}=m \ne 0,$ $\,\forall  i$.  It was found in  \cite{CKM}  that each of our confining  singularities   of the $USp(2N)$  theory  (Chebyshev vacua)   splits
  up into  various singularities  describing  local   $SU(r)\times U(1)^{N-r}$  theories, $r=0,1,\ldots, \frac{N_{f}}{2}$.  Physics in each of these $r$ vacua is  similar as in the
  ``r-vacua''  in the $SU(N)$ QCD, due to the universality of SCFT,  as pointed out  by Eguchi et. al. \cite{Eguchi}.     We know then, {\it assuming} that the light  chiral
  multiplets present for $m_{i}=m \ne 0$ survive in the $m \to 0$ limit,   that there are  monopoles and dyons  with  various
 effective   $SU(r)$  charges, and in the fundamental representation of the global $SU(N_{f}) \subset SO(2 N_{f})$   group.

 We do know that in the SCFT limit ($m \to 0$ limit), these monopoles all become massless simultaneously, but they are relatively non local, carrying mutually non zero  Dirac units \cite{TM}.  Local subset of fields, belonging to one of the $r$ vacua, realize only a subgroup
 $SU(N_{f})$ of the full symmetry group $SO(2 N_{f})$.  On the other hand the Seiberg-Witten curve of this  theory
 \be      x y^{2} = \left[ x \prod_{a=1}^{N} (x-\phi_{a}^{2})
          + 2\Lambda^{2 N+2-N_{f}} m_{1} \cdots m_{N_{f}} \right]^{2}
          - 4\Lambda^{2(2 N+2-N_{f})} \prod_{i=1}^{N_{f}}(x+m_{i}^{2})
      \label{eq:curveusp}
 \ee
 has the correct flavor symmetry structure, $SO(2N_{f}) $ in the $m_{i}\to 0$ limit  (see for instance \cite{AS}).

 We have,  apparently, an interesting new physical situation in which the global symmetry ($SO(2 N_{f})$)  of the system is realized by fields,  mutually local subsets of which realize only a subgroup of the full symmetry group. This is perhaps not really surprising, once one  accepts the fact that
 the system under study has  no local low-energy  effective action description.

 As pointed out  already, it is not easy to compute anything explicitly in these circumstances,  but it is reasonable to assume that the symmetry breaking  (Eq.~(\ref{similar})) is induced  by the condensates of the monopoles $ M_{a}^{i}$ and   $ {\tilde M}_{j}^{a}$,  with $a=1,2,\ldots, \left[\frac{N_{f}}{2}\right]  $,
\be   \brc  \, M_{a}^{i}   {\tilde M}_{j}^{a} \ckt  = v\,  \delta_{j}^{i} \ne 0,  \qquad  i,j =1,2, \ldots, N_{f}. \label{monocondUSpN}
\ee
Note that these monopoles  are the most strongly interacting fields, with $SU(\left[\frac{N_{f}}{2}\right])$ gauge interactions;
other  monopoles carrying $SU(r)$  charges  $r < \left[\frac{N_{f}}{2}\right]$,  are more weakly coupled in the infrared \footnote{
Actually in the limit $m \to 0$, the full gauge symmetry is recovered;   it is not easy to see  the effects of the  $USp(2N)$ interactions among  the nonlocal sets of monopoles.  The situation of the equal mass limit discussed in Section
\ref{SUNSCFT} was however  different:  there,  the strong interactions among the monopoles arise only in the  SCFT limit.
}.

Eq.~(\ref{monocondUSpN})  also naturally generalizes the result of the $USp(4)$  case, Eq.~(\ref{monocond}).
Again, we do not expect baryonlike condensate (analogue  of (\ref{condensgen}))  to form in this case, in contrast to what happens in $SU(N)$
theories.

This is probably related  to the absence of gauge-invariant
baryonlike condensate (\ref{condensgen}) in the $USp(2N)$  theory, which is
instead present and crucial for $SU(N)$  theories.   This is a
manifestation of the fact that the  SCFT under consideration in the
$USp(2N)$ theory is in a different   universality class from  that
of the  $r= \left[\frac{N_{f}}{2}\right]$ vacuum in $SU(N)$ theory
(as argued in \cite{CKM}): they have different   global symmetries,
and light degrees of freedom and their interactions are distinct.

To understand better the behavior of the monopoles and to see the
posssible relevance of Goddard-Nuyts-Olive (GNO) monopoles
\cite{GNO},
we looked into the perturbation (by masses $m_{i}$) around the SCFT
points,  studied in \cite{CKM}, a little more carefully. For
definiteness, let us  consider the Chebyshev point of  $USp(2 N )$
theory  with $N_f =  2 \, N +1 $ and, instead of keeping
nonvanishing equal masses and sending all of them to zero as done above,  we first keep one mass much larger, and send others to
$0$ first, {\it i.e.,}    $\Lambda \gg m \gg m_i\to 0$,
$i=2,3,\ldots, N_f$.   If $m$ were large, $m \gg \Lambda$,  the
gauge group will be broken  to  $USp(2 N -2  ) \times U(1)$  and  a
semiclassical reasoning will tell us that  the system  contains
massive nonabelian monopoles, transforming in the dual gauge group
$SO(2N-1)$, see \cite{NAmonop}-\cite{ ABEKM}. The number of the
flavor ($N_f =  2 \, N +1 $) is chosen such that the ``unbroken''
group $USp(2 N -2  )$ is infrared free, so that it does not break
itself further dynamically. We are particularly interested in
knowing whether such monopoles survive quantum effects and  become
light, as $m$ is reduced to smaller values,  {\it i.e.},  less than
$\Lambda$.  The behavior of the theory in such a limit  is encoded
in the Seiberg-Witten curve, Eq.~(\ref{eq:curveusp}). The curve
reduces at  the mass scales much lower than  $m$  (where $x  \ll m$,
) to the form (see \ref{appA}):
\be   x \, y^2 \simeq   4 \, m^{2} \, \Lambda^2 \, \left[   \left( x\,\prod_{i=2}^{N}  (x - \phi_i^2)  -  \prod_{i=2}^{N_f} m_i \right)^2
-    \prod_{i=2}^{N_{f}}   (x + m_i^2) \right], \label {questo}
\ee
which is nothing but the curve of  $USp(2 N^{\prime})=  USp(2 N - 2
)$ theory with  $N_{f}^{\prime }= N_f-1 =  2 N = 2 N^{\prime}+ 2$.
By comparing this curve with the general form of the curve given in
\cite{AS},  one sees that it is a SCFT  (in the limit $m_{i} \to 0$)
  with infinitely strong coupling\footnote{   (\ref{questo}) coincides with the curve given in \cite{AS}  with  $g(\tau) =   -1$, where
  \[  g(\tau) \equiv  \frac{\vartheta_{2}^{4}(\tau) }{ \vartheta_{3}^{4}(\tau) +  \vartheta_{3}^{4}(\tau)}, \qquad  \vartheta_{2}(\tau)=\sum_{n \in {\mathbb Z}} q^{(n + 1/2)^{2}};\quad  \vartheta_{3}(\tau)=\sum_{n \in {\mathbb Z}} q^{n^{2}};\quad \vartheta_{4}(\tau)=\sum_{n \in {\mathbb Z}} (-)^{n} \, q^{n^{2}}.
\]
}.     No hint of GNO monopoles, which would transform according to the dual gauge group $SO(2N -1)$, is there.

If now another mass ({\it e.g.,}  $m_{2}=m^{\prime}$) is kept fixed and other masses ($i=3,4,\ldots$)
are sent to zero first, the theory below   the scale $m, m^{\prime}$  will behave as an $USp(2N-2)$ theory with one flavor less, an asymptotically free theory. But as $g(\tau) = - 1$ and the effective RG invariant scale is $\Lambda^{\prime}  = g(\tau) \, m^{\prime}  \sim   m^{\prime}$, the RG invariant scale of the theory is large. In other words the low energy
theory is a strogly coupled system, and cannot be described by a weakly-coupled picture of monopoles.

An analogous result holds for even $N_{f}$, for instance, $N_{f}= 2 N$:   the theory in the regime
 $\Lambda \gg m \gg m_i\to 0$,   $i=2,3,\ldots, N_f$  is a strongly interacting system.  The moral of the story is that there are no weakly-coupled
 monopoles acting as infrared degrees of freedom in the confining  vacuum around the SCFT  of the softly broken ${\cal N}=2$  $USp(2N)$
 theory and, in particular, there are no hint of GNO monopoles becoming light by quantum effects in these theories.

 Analogous conclusions apply also to  the Chebyshev (SCFT) vacua of $SO(N)$  theories.

\section{Abelian versus nonabelian monopoles  \label{disaster}}

   As is well known,  't Hooft-Polyakov monopoles can acquire, upon quantization of matter fields, flavor quantum (quark) numbers through the fermion zero modes \cite{JR}. If such a  monopole  condenses, inducing confinement, the global symmetry is also broken  dynamically,    leading to  a direct connection between confinement and dynamical symmetry breaking \cite{SW2}.

As was noted in \cite{CKM},  however, in certain vacua  there are reasons to believe that  abelian monopoles cannot possibly  be  the correct infrared degrees of freedom. For concreteness, let us first  consider the case of ${\cal N}=2$ supersymmetric
 $USp(2N)$  gauge theories with $N_{f}$ fundamental matter multiplets, discussed in the preceding section,  where   the global symmetry is $SO(2N_{f})$.
  According to the standard  Jackiw-Rebbi analysis \cite{JR}   the semiclassical 't Hooft-Polyakov monopoles of this theory  are in an even or  odd spinor representations of $SO(2N_{f})$, each with multiplicity, $2^{N_{f}-1}$.
  The number of ${\cal N}=1$ vacua fits very nicely with it \cite{CKM}.

  However, if the abelian monopoles were the light degrees of freedom  the low-energy  theory would have automatically a global,  accidental
$SU(2^{N_{f}-1})$   symmetry, and would lead eventually to a huge number of Nambu-Goldstone bosons, not expected from the symmetry considerations alone.

Curiously, in the $SU(2)$ theories studied by Seiberg and Witten \cite{SW2}, this does not cause a problem, apparently simply because
  $2^{N_{f}-1}$ (for $N_{c}=2$, $N_{f}$ is limited by $N_{f}\le 3$)   is  a small number!  The light monopoles {\it are} found to be  indeed 't Hooft-Polyakov abelian monopoles in the spinor representations of  $SO(4)\sim SU(2) \times SU(2)$ and of $SO(6) \sim SU(4)$ for $N_{f}=2$ and $N_{f}=3$, respectively \cite{SW2}.  Monopole condensation leads to the symmetry breaking
\be   SO(4) \to  SU(2), \qquad {\rm or} \qquad  SO(6) \to SU(3),
\ee
with ensuing massless spectrum compatible with the standard Nambu-Goldstone theorem.

Once the rank of the gauge group is taken higher, one immediately faces a question:
could it be possible that the effective low-energy theory possesses a global symmetry which is much
larger than the true symmetry of the system, {\it i.e.}, a very large accidental symmetry?
Is it possible that the low-energy spectrum  contains  many  massless particles,  expected neither
from the Nambu-Goldstone theorem nor from any other principles (such as 't Hooft's anomaly matching condition)?

Actually, none of these embarrassing things  occur.  In the $USp(2N)$ theory with matter,
the potential abelian mono\-poles  in  the spinor representation of $SO(2N_{f})$,  actually  are replaced by  the nonabelian  monopoles  in the fundamentals of various $SU(r)$
effective gauge groups, and transforming as ${\underline {N_{f}}}$ or  ${\underline {N_{f}}^{*}}$    of the subgroup $SU(N_{f}) \subset SO(2N_{f})$.

 \smallskip

      The replacement of the potential 't Hooft-Polyakov monopoles by  nonabelian mono\-poles, is not a phenomenon
restricted  to the Argyres-Douglas vacua where monopoles are strongly coupled: it
  occurs also in an $r$ vacuum of the  $SU(N)$ gauge theory, as was already noted in \cite{CKM}.  Physics of the  $r$ vauca in that theory is quite well understood, except the SCFT case $r= \frac{N_{f}}{2}$ discussed in Section \ref{SUNSCFT}:  it is an effective $SU(r)$ gauge theory with $N_{f}$ dual quarks (nonabelian Goddard-Nuyts-Olive monopoles \cite{GNO}), and  confinement is induced by the condensation of these nonabelian monopoles. As long as the flavor quantum numbers are concerned, it looks  as if  abelian  monopoles  in the antisymmetric rank $r$  tensor representation of the global $SU(N_{f})$ symmetry in these vacua had broken  up \cite{CKM}   into ``baryonic'' components, as
 \be      M_{tHP}^{i_{1}\ldots i_{r}}  \sim     \epsilon^{a_1   \ldots  a_r} {q_{a_1}^{i_1} q_{a_2}^{i_2}
\ldots q_{a_r}^{i_r} }.
 \ee
 The nonabelian monopoles ($q_{a}^{i}$'s)  carry  $\frac{1}{r}$ of the $U(1)$ magnetic  charge with respect to what is expected for a  minimal {\it abelian} monopole.   This is  understood by the fact that the nontrivial homotopy group
 \be      \pi_{1}(U(r)) = \pi_{1}\left(\frac{SU(r)\times U(1) }{{\mathbb Z}_{r}} \right)  =  {\mathbb   Z}
 \ee
 is generated by the minimal loop involving the ${\mathbb Z}_{r}$ element of $SU(r)$, that is, a $\frac{1}{r}$ of the full circle in the $U(1)$
 gauge factor (see for instance  the first of \cite{ABEKM}).

       Let us add a remark  that the dual quarks in the Seiberg's dual theory in $SU(N)$   SQCD, can also be regarded as the ``baryonic''
       components (with respect to the dual $SU({\tilde N})$ group) of the original baryonic composites
       $B = \epsilon_{a_{1}\ldots a_{N}}   Q^{a_{1}}  Q^{a_{2}}  \ldots Q^{a_{N}}  $ of the fundamental theory:
       \be    B \sim  \epsilon_{b_{1}\ldots b_{\tilde N}}   q^{b_{1}}  q^{b_{2}}  \ldots q^{b_{\tilde N}}, \qquad {\tilde N}=  N_{f}-N.
       \ee
        In this case it is the constraint of the 't Hooft's anomaly matching (correct low-energy symmetry realization)  that forces the system to choose Seiberg's dual quark as the infrared degrees of freedom.

\section{Quantum vs classical  $r$-vacua of $SU(N)$  theories:\\
Seiberg's dual quarks as GNO monopoles}

What is the  physical meaning of Seiberg's duality  in ${\cal N}=1$ supersymmetric QCD?
In spite of the observation made at the end of the preceding section, and in spite of
overwhelming evidence for it \cite{Sei},  the nature  of the  Seiberg's duals remains somewhat mysterious.
Attempt to ``understand''  it starting from ${\cal N}=2$ supersymmetric QCD \cite{APS} was not entirely successful.

A   clue to  the meaning of the Seiberg's dual  quarks comes from  the
exact  relation between the quantum and classical $r$-vacua appearing in the
${\cal N}=2$ supersymmetric $SU(N)$ QCD \cite{BKM}.  For definiteness and for concreteness let us restrict ourselves to the
cases of $SU(N)$ gauge groups in this section.

 In Carlino et. al. \cite{CKM}
the matching of the {\it total number}  of semiclassical (at large $\mu$ and $m_{i}$) and quantum mechanical (at small  $\mu$ and $m_{i}$)   ${\cal N}=1$  vacua  has been worked out carefully.  Although not stated very explicitly there, the counting and matching work out  for  the vacua having a definite symmetry, actually:  we are
able to ``follow'' the flow of each vacuum from the semiclassical region to fully quantum mechanical regime \footnote{As in Carlino et. al. \cite{CKM},
we keep nonvanishing and generic masses $\mu$ and $m_{i}$ so that only discrete set of ${\cal N}=1$ vacua remain. }.
First consider the case $N_{f} \le N$.  It is seen from (\ref{globalSB}) and (\ref{clasbr})   that there are  total of
\be \sum_{r=0}^{N_{f}} (N-r) {N_{f} \choose {r}}  =  ( 2 N - N_{f}) \, 2^{N_{f}-1}
\label{classical}\ee
vacua, where $r$  is the number of the massless flavors in the semiclassical theory.  In fact,  the index $r$ chracterizes both the local gauge symmetry
($SU(r) \times U(1)^{N-r}$)  and the flavor symmetry ($U(r) \times U(N_{f}-r)$)  of each vacuum.

In the counting of classical vacua (\ref{classical})  the integer $r$ runs from $0$ to $N_{f}.$   The number of the vacua with  global  $U(r_{0})\times
U(N_{f}-r_{0})$  symmetry ($r_{0} < \frac{N_{f}}{2}$)   is given by the sum of those with $r=r_{0}$ and   $r=N_{f}-r_{0}$,
\be   (N - r_{0})    {N_{f} \choose {r_{0}}} +    \{ N - ( N_{f} -r_{0}) \}    {N_{f} \choose {N_{f}- r_{0}}}  =  (2 N- N_{f}) \,  {N_{f} \choose {r_{0}}},
\ee
 which precisely matches the number of the quantum $r$ vacua with  $r=r_{0}$.  Note that  in the quantum case
 $2 N- N_{f} $ is  the multiplicity due to the discrete  ${\mathbb Z}_{2 N- N_{f}}$  symmetry present in the $m_{i} \to 0$ limit;   $r_{0} $ runs only up to $\frac{N_{f}}{2}$.  This last fact can be understood on the basis of the renormalization group argument \cite{BK}.      Summarizing, the classical and quantum $r$ vacua correspond as
 \be   \{ r_{0},  N_{f} - r_{0} \}  \to  r_{0}, \qquad   r_{0} \le \frac{N_{f}}{2}.
 \ee

 This analysis   is useful, not so much as a further consistency check   but because it tells us  something nontrivial about physics.
 Semiclassically the $SU(r)$ gauge theory with $r=r_{0}<  \frac{N_{f}}{2}$ is infrared free \footnote{We recall that the beta function of the ${\cal N}=2$ SQCD is proportional to  $N_{f}- 2 N_{c}$. }, 
  so it is natural that it survives in the infrared, even though  the quarks ``become'' monopoles  due to the rearrangement of singularities (phenomenon of ``isomonodromy'' discussed in \cite{SW2,BF,Cap}).


 The theory with classical $SU(r)$ gauge symmetry with  $r= N_{f}-r_{0} >\frac{N_{f}}{2}$, instead, is asymptotically free and becomes strongly coupled in the infrared.
 The vacuum counting and matching above teach us that {\it  this theory is replaced in the infrared by an
 $SU(N_{f} -r)  = SU(r_{0})$  theory with $N_{f}$ flavors of nonabelian monopoles.}   This is nothing but  the Seiberg's duality \cite{Sei}.

 The important point is that the monopoles in these vacua can be  identified  \cite{BK}  at the same time also  as  the nonabelian Goddard-Nuyts-Olive monopoles \cite{GNO}
 associated with the  partial  gauge symmetry breaking
\be  SU(N)  \to  SU(r_{0}) \times  U(1)^{N-r_0},   \qquad  r_{0} < \frac{N_{f}}{2} \le N,
\ee
since the fundamental theory contains  the adjoint scalar $\Phi$.   This provides a precious bridge between the concept of semiclassical nonabelian
monopoles   and that of  Seiberg's dual quarks.

And this hints at the interpretation of Seiberg's dual quarks as the GNO monopoles also in the standard, ${\cal N}=1$ supersymmetric QCD.
Let us recall that for $N+1  <     N_{f} \le 3 N $  the concept of magnetic quarks make sense, the dual theory is a $SU({\tilde N}) =  SU(N_{f}- N)$  theory,  while
for $\frac{3 N}{2} \le  N_{f} $ the concept  of the quark field make sense as the infrared degrees of freedom.   In the conformal window,
  \be     \frac{3 N}{2} \le  N_{f} \le 3 N
  \ee
either description can be used, and the theory flows into an infrared fixed-point theory (superconformal theory).  Now,
when $N_{f} <  2 N$  the original $SU(N)$  theory is more strongly coupled in the infrared: it is natural to consider it as the ``fundamental'' theory and
$SU({\tilde N})$ as its dual  (${\tilde N} < N$);
for $N_{f} >   2 N$  the   $SU({\tilde N})$  theory might be considered as   fundamental,  the quarks are then GNO ``monopoles''.
Of course, the   ${\cal N}=1$ SQCD  does not have an elementary  scalar field in the adjoint representation,
 and there are  no classical soliton monopoles.   Nevertheless the theory seems to
produce dynamically magnetic soliton-like monopoles, which act as the correct infrared degrees of freedom.
And this, in turn, seems to suggest that an analogous phenomenon is  possible in the ordinary QCD.

 The correspondence between classical and quantum vacua are  subtler in the case, $N_{f} > N$.
 One must distinguish the  cases with  $r \ge  N_{f}-N$ and those with $r <  N_{f}-N.$   For the vacua $r \ge  N_{f}-N$
 the discussion analogous to the one done for $N_{f} \le  N$ holds,  there are a pair of classical vacua  $(r_{0},  N_{f}-r_{0})$   in which the symmetry
 of the system is $U(r_{0})\times U(N_{f}- r_{0})$.  Their sum give
 \be  \sum_{r_{0}= N_{f}-N}^{N}  (N- r_{0}) {N_{f}  \choose {r_{0}}} =  \sum_{r_{0}= N_{f}-N}^{N_{f}/2}  (2 N-  N_{f}) {N_{f}  \choose {r_{0}}}.
 \label{firstgr}\ee

 The vacua with smaller $r=r_{0}$, $r_{0} <  N_{f}-N$  appear alone: their total number is
 \be   \sum_{r_{0}=0}^{N_{f}-N-1}  \,  (N - r_{0})    {N_{f} \choose {r_{0}}}
\label{alone} \ee
Quantum mechanically, we know that there are two kinds of vacua,
  those in the confinement phase appear  \be {\cal N}_{1}=  \sum_{r_{0}=0}^{N_{f}/2} (2 N- N_{f}) \,  {N_{f} \choose {r_{0} } } \ee  times,
  while the number of the vacua in a free magnetic phase (no confinement)  is \be {\cal N}_{2}=   \sum_{r_{0}=0}^{N_{f}-N-1}  \,  (N_{f} -  N  - r_{0})    {N_{f} \choose {r_{0}}}.
 \ee
In order to find the matching, split the sum   in (\ref{firstgr})  as
 \be    \sum_{r_{0}= N_{f}-N}^{N_{f}/2}  (2 N-  N_{f}) {N_{f}  \choose {r_{0}} }  =   \sum_{r_{0}=0}^{N_{f}/2}   (2 N-  N_{f}) {N_{f}  \choose {r_{0}} }   -  \sum_{r_{0}= 0}^{N_{f}-N -1 }  (2 N-  N_{f}) {N_{f}  \choose {r_{0}} }.
\label{splitting}\ee
 The first term is equal to  ${\cal N}_{1}$,  while the sum of the second term  with (\ref{alone})
 is precisely ${\cal N}_{2}$.

\section {GNO monopoles in $USp(2N)$,   $SO(2N+1)$ theories}

We have presented  in Section \ref{lightmpl} some evidence against
the appearance of light GNO monopoles associated with partial
$USp(2N)$ gauge group breaking in theories with ${\cal N}=2$
supersymmetry. Actually there is a simple reason why light monopoles
of GNO type cannot appear as light degrees of freedom, at least in
the context of  ${\cal N}=2$  supersymmetric theories, with
$USp(2N)$ or $SO(N)$ groups.  Suppose that the gauge symmetry is
partially broken, for instance, as
\be  USp(2N) \to  USp(2N-2) \times U(1).
\ee
The semiclassical monopoles representing the homotopy group
\be   \pi_{2}\left( \frac{USp(2N) }{USp(2N-2) \times U(1)} \right) \sim \pi_{1} ({USp(2N-2) \times U(1)})
\ee
would transform as in the fundamental representation of $ SO(2N-1)
\times U(1)$,  dual of $USp(2N-2) \times U(1)$. Now in the presence
of $N_{f}$ massless  flavors (which is needed to prevent $USp(2N-2)
$ from breaking itself dynamically to abelian subgroups), the
fundamental theory has a global  $SO(2N_{f})$ symmetry, as already
mentioned. Now if the GNO  monopoles appeared in the low energies,
the low-energy effective theory would have an $USp(2N_{f})$
symmetry, instead of the correct $SO(2N_{f})$ symmetry \footnote{We
thank H. Murayama on the discussion on this point.}. The global
symmetry appears to prevent the GNO monopoles from becoming  light
in these circumstances, and indeed this does not occur!

An analogous situation presents itself in $SO(2N +1)$  gauge
theories with $N_{f}$ flavors in vector representation,  where the
global symmetry is $USp(2N_{f})$. The GNO monopoles associated with
partial breaking , {\it e.g.}, $SO(2N +1)\to SO(2N-1) \times U(1)$,
cannot become light, as  it would imply a wrong symmetry in the low
energies, and indeed,  they do not appear.

\section {Discussion}

In this paper we have discussed various aspects of nonabelian
Argyres-Douglas vacua which occur frequently in supersymmetric
theories. The main observation  made here  is that,  besides the
renormalization group features  already  emphasized  in
\cite{BK,ABEKM}, there are other constraints, mostly related to the
realization of the global symmetries,    which severely restrict
which types of monopoles can appear as  the light degrees of
freedom. We have analysed different classes of $SU(N)$ and $USp(2N)$
theories  (mentioning also  briefly results on $SO(N)$  theories) 
and found a number of interesting phenomena. Although the fact that
we are able to analyse and get information does depend on the
presence of supersymmetry, we believe that the phenomena themselves  are of much 
more general nature.

The most interesting, and quite general,  phenomenon is the replacement of
abelian monopoles by nonabelian monopoles as the infrared degrees of freedom.
 We found various cases in which the system is forced to  choose the latter, in order to
 realize appropriately the global symmetry of the underlying theory, {\it e.g.}, not to have too large an accidental symmetry.

Secondly, in a wide class of nonabelian Argyres-Douglas vacua studied here,  the confinement (which requires a partial supersymmetry breaking)  is  induced  by the condensation of strongly-interacting nonabelian monopole composites.  These monopoles carry the
 flavor charges, and  the pattern of global symmetry breaking is  determined  by the type  of {\it composites } of nonabelian monopoles which condense,  such as (\ref{condens}), (\ref{condensgen}),   (\ref{monocond}) and (\ref{monocondUSpN}).

Thirdly -  this seems to be the most important feature valid in wide classes of supersymmetric models -  the most interesting confining vacua with nonabelian symmetry,
namely, confinement neither accompanied by dynamical abelianization nor by infrared-free dual gauge interactions,  occurs near ({\it i.e.,} perturbation of) nontrivial conformal vacua. 

What can one learn from these studies for QCD?   Let us speculate on what might be  possibly happening, drawing analogies from some of the phenomena found here.
In the standard QCD, the lattice calculations tell us that the chiral symmetry restoration and deconfinement transitions  take place   at the same temperature, suggesting a close dynamical connection between the two phenomena.
 There are no hint of dynamical abelianization, so that if  the ground state of QCD is a kind of dual superconductor,
 it must be of a nonabelian variety \cite{ABEKY}.  The $SU(3)$ color group is not broken dynamically to $U(1) \times U(1)$; let us assume that the dual theory is instead $SU(2)\times U(1)$.   The monopoles  $M_{\alpha}^{i}$ and    ${\tilde M}^{\alpha}_{j}$
 would carry the dual color index  ${\alpha =1,2}$  and the flavor indices   $i$ (of $SU_{L}(N_{f})$)  and  $j$ (of $SU_{R}(N_{f})$), respectively.    As there are no  phenomenological evidences that the long-distance hadronic physics is governed  by weakly-coupled magnetic monopoles, we must assume that they interact strongly.

 We speculate that the condensate
 \be   \brc  \, M_{\alpha}^{i}   {\tilde M}_{j}^{\alpha} \ckt  = v\,  \delta_{j}^{i} \ne 0,  \qquad  i,j =1,2, \ldots, N_{f}, \qquad
 v \sim O(\Lambda_{QCD}^{2} )   \label{monocondQCD}
\ee
is formed, due to the strong dual gauge interactions, inducing confinement, and at the same time causing the symmetry breaking
 \be  SU_{L}(N_{f}) \times  SU_{R}(N_{f})  \times U_{A}(1) \times U_{V}(1) \to SU_{V}(N_{f}) \times  U_{V}(1). \label{SBQCD}
 \ee
One might wonder why the condensing entity cannot simply be a 't Hooft-Polyakov (or 't Hooft-Mandelstam) monopole, dressed with  flavor quantum numbers of bifermion composite, ${\bar \psi}_{R  } \, \psi_{L}$.  It  would however  be  inconsistent to assume that such abelian monopoles are the dominant effective degrees of freedom.  A $U(1)$  theory with $N_{f}^{2}$ monopoles  would have a global symmetry, $SU(N_{f}^{2})$,  and the disaster  of  ``too-many-Nambu-Goldstone bosons''  discussed  in Section \ref{disaster} would ensue.  As the supersymmetric systems discussed there, the standard $SU(3)$ QCD would probably avoid such an awkward  situation by producing  nonabelian monopoles (even at the price of having them  necessarily strongly interacting)
 as the effective degrees of freedom.

The condensate  (\ref{monocondQCD})  carries  the same flavor quantum numbers as the standard  quark bilinear condensate,
$ \brc {\bar \psi}_{R \, j } \,  \psi_{L}^{i} \ckt$.
 It is quite possible that these two types of condensates are actually dynamically  related, either by a nonabelian Jackiw-Rebbi mechanism (through the fermion  zeromodes), or by a generalization of the  Rubakov effect \cite{Rubakov}.


A somewhat related idea, that the vacuum of QCD is close to a nontrivial infrared fixed-point theory and that such a conformal invariance is achieved by the collaboration of relatively nonlocal fields, has been discussed recently by C.R. Das et. al. \cite{Nielsen}.

\section*{Acknowlegdment}

The authors thank Roberto Auzzi, Stefano Bolognesi, Nick Dorey, Jarah Evslin, Roberto Grena  and Hitoshi Murayama, for useful discussions
at various stages of the present  work.

\appendix

\section {Light monopoles at  the Chebyshev point of $USp(2N)$  theory
 \label{appA}}

The curve of the theory is given by
 \be      x y^{2} = \left[ x \prod_{a=1}^{N} (x-\phi_{a}^{2})
          + 2\Lambda^{2N+2-N_{f}} m_{1} \cdots m_{N_{f}} \right]^{2}
          - 4\Lambda^{2(2N+2-N_{f})} \prod_{i=1}^{N_{f}}(x+m_{i}^{2}).
      \label{Curveusp}
 \ee
 For definiteness first consider the case of odd number of flavors,  with $N_{f}= 2 N +1.$
The Chebyshev solution corresponds to
 $\phi_{a}=0,$   $\forall a$:
\begin{equation}
      x y^{2} = \left[ x^{N+1} \right]^{2}
          - 4\Lambda^{2} x^{2N+1}
          = x^{2N+1}(x-4\Lambda^{2}).
\end{equation}
The zero at $x=0$ is of degree $2 N$, and there is another
isolated zero at $x=4\Lambda^{2}$.  There is also a branch point at
$x=\infty$.

Under the perturbation by generic quark masses, the problem is to find $\{\phi\}$'s such that
 the curve  factorizes with maximal number of  double factors as
\begin{equation}
      x y^{2} = x (x-4\Lambda^{2} -\beta) \prod_{a=1}^{ N}
      (x-\alpha_{a})^{2},
\end{equation}
and to find out in how many ways this can be done. This problem has been solved in Section 9.2.
of \cite{CKM};  the answer is
\begin{equation}
      s_{k}(\alpha) = s_{2k}(m), \qquad
      s_{k+1}(\phi^{2}) = - 2\Lambda (-1)^{ N} s_{2k+1}(m), \quad k=0,1,2, \ldots,
      \label{eq:solutions}
\end{equation}
where the  symmetric polynomials $s_{j}(\rho)$ are defined by:
\begin{equation}
      \prod_{i=1}^{N} (z-\rho_{i})
      = \sum_{k=0}^{N} (-1)^{k} s_{k}(\rho) z^{N-k},
\end{equation}
 that is,  $s_{0}(\rho) =
1$, $s_{1}(\rho) = \sum_{i=1}^{N} \rho_{i}$, $s_{2}(\rho) =
\sum_{i<j} \rho_{i} \rho_{j}$, etc.

We now consider a particular set of masses, $m \gg m_i$,   $i=2,3,\ldots, N_f$.
From Eq.~(\ref{eq:solutions})  it follows that
\[ s_1(\phi^{2})  =  \sum_{i=1}^{N}  \, \phi_i^2
= 2 \, \Lambda \, \sum m_i \simeq  2 \, \Lambda\, m; \]
\[ s_2(\phi^{2})  =  \sum_{i<j} \, \phi_i^2\, \phi_j^2  \simeq \phi_1^2 \, s_1^{\prime}(\phi^{2})
= 2 \, \Lambda \, s_3(m_i)  \simeq  2 \, \Lambda\, m \, s_2^{\prime}(m_i); \]
\[ s_3(\phi^{2})   \simeq \phi_1^2 \, s_2^{\prime}(\phi^{2})
= 2 \, \Lambda\, m \, s_4^{\prime}(m_i); \]
etc., and in general
\[   s_k^{\prime}(\phi^{2})  \simeq s_{2 k}^{\prime}(m_i). \]
As for $\alpha$, one finds
\[ s_1(\alpha) \simeq m \,  s_1^{\prime}(m_i); \qquad   s_2(\alpha) \simeq m \,  s_3^{\prime}(m_i);\qquad   s_1(\alpha) \simeq m \,  s_5^{\prime}(m_i),\] and so on, and in general
\be  \alpha_1 \simeq  m \, s_1^{\prime}(m_i), \qquad  s_k^{\prime}(\alpha)\simeq  \frac{s_{2k + 1}^{\prime}(m_i)}{s_{1}^{\prime}(m_i)}.
 \ee
 The prime above means
  $\phi_1^2$, $\alpha_{1}$ and $m_1$  do not appear.
  The above results suggest that
  \be \phi_{1}^{2}\sim 2 \, m \, \Lambda, \qquad  \phi_{i}^{2} = O(m_{j}^{2}), \quad i=2,3,\ldots.\ee
The curve then looks like (where $x  \ll m$)
\be   x \, y^2 \simeq   4 \, m^{2} \, \Lambda^2 \, \left[   \left( x\,\prod_{a=2}^{N}  (x - \phi_a^2)  -  \prod_{i=2}^{N_f} m_i \right)^2
-    \prod_{i=2}^{N_f}   (x + m_i^2) \right], \ee
which is nothing but the curve of  $USp(2 N^{\prime})=USp(2 N - 2 )$ theory with   $N_f^{\prime }= N_f-1 =  2 N  = 2 N^{\prime}+ 2$.  It is a SCFT  with     $g(\tau)=-1$ (infinitely strong coupling) \cite{AS}, where
  \be  g(\tau) =
          \frac{\vartheta_{2}^{4}}{\vartheta_{3}^{4}+\vartheta_{4}^{4}}.
\ee
If now another mass ({\it e.g.,}  $m_{2}=m^{\prime}$) is kept fixed and other masses
are sent to zero, the theory below   the scale $m, m^{\prime}$  will behave as an $USp(2 N - 2)$ theory with one flavor less, an asymptotically free theory. But as $g(\tau) =- 1$ and the effective RG invariant scale is $\Lambda^{\prime}  = g(\tau) \, m^{\prime}  \sim  m^{\prime}$, the RG invariant scale of the theory   is large. In other words the low energy
theory cannot be described by a weakly coupled theory.

The Chebyshev point of  $USp(2N)$  theory  with  even number of flavors,  $N_f =  2 \, N $,  with one mass kept finite,   $\Lambda \gg m \gg m_i\to 0$,   $i=2,3,\ldots, N_f$, can be treated in a similar manner.
We first note that at $\Lambda \gg m_{i}$,  $\forall i$,   the curve effectively reduces \cite{CKM}  to
  \begin{equation}
      x y^{2} \simeq  (2\, \Lambda^{2})^{2}  \left[ x \prod_{a=1}^{N-1} (x-\phi_{a}^{2})
          -   m_{1} \cdots m_{N_f} \right]^{2}
          -    \prod_{i=1}^{N_f}(x+m_{i}^{2}),
\label{arrangedbis}  \end{equation}
which is an SCFT  with $g =-1$,   as
\be  N^{\prime} = N-1, \qquad  N_f = 2\, N =   2\, N^{\prime} + 2.
\ee
 Again, by choosing one of the bare masses to be large as compared to the others,
$m_{1} \gg m_i $,  $i=2,3, \ldots $,  one gets effectively an asymptotically free theory with
\be  N^{\prime}= N-1, \qquad  N_f^{\prime} = 2\, N^{\prime} +1, \ee
and  with the effective RG invariant scale,
\be \Lambda^{\prime} = m_{1},
\ee
which is large at the mass scale $ m_{i} \ll  m_{1}$. The theory is strongly coupled.

Analogous conclusions apply to   the Chebyshev (SCFT) vacua of $SO(N)$  theories with massless matter, as can be shown
by using the results of \cite{CKKM}.

\end{document}

The underlying action is that of  ${\cal N}=2$ SQCD   with superpotential,
\be   {\cal L} =  \frac{1}{8\, \pi} \Im \, \tau \,\left[ \int   d^{4}\theta\, \Phi^{\dagger}\, e^{V}\, \Phi
+ \int d^{2} \theta \frac{1}{2} \, W^{\alpha}\, W_{\alpha} \,\right]  + {\cal L}^{(quarks)} + \int d^{2} \theta {\cal W} (\Phi, Q,{\tilde Q}) \ee
where $m_{i}$ are the ``bare''  quark masses,
\be{\cal L}^{(quarks)}= \sum_i \, [ \int d^4 \theta \, \{ Q_i^{\dagger} e^V
Q_i + {\tilde Q_i}^{\dagger}  e^{ {\tilde V}}    {\tilde Q}_i \} +
\int d^2 \theta
\, \{ \sqrt{2} {\tilde Q}_i \Phi Q^i    +      m_i   {\tilde Q}_i    Q^i   \}
\label{lagquark}
\ee
and  the ${\cal N}=1$ superpotential is given by
\be  {\cal W} (\Phi, Q,{\tilde Q})   =     \int \, d^2 \theta \,\mu  \,\Tr \, \Phi^2   \ee
gives the mass to the adjoint scalar multiplet and
reduces the supersymmetry to $N=1$.
\be
\tau \equiv  {\theta_0 \o \pi} + {8 \pi i \o g_0^2}
\label{struc}
\ee

\end{document}